\documentclass[10pt,english]{article}
\usepackage{times}
\usepackage[T1]{fontenc}
\usepackage[latin1]{inputenc}
\usepackage{amsmath}
\usepackage{amssymb}

\makeatletter

\usepackage{slashed}

\usepackage{babel}
\makeatother
\begin{document}

\title{\textbf{THE ELECTRIC CHARGE ASSIGNMENT IN $SU(4)_{L}\otimes U(1)_{Y}$
GAUGE MODELS}}

\author{ADRIAN PALCU}

\date{\emph{Faculty of Exact Sciences - {}``Aurel Vlaicu'' University
Arad, Str. Elena Dr\u{a}goi 2, Arad - 310330, Romania}}

\maketitle
\begin{abstract}
In this brief report, apart from the usual approach, we discriminate
among models in the class of \textbf{$SU(4)_{L}\otimes U(1)_{Y}$}
electro-weak gauge models by just setting the versors in the method
of the exactly solving gauge models with high symmetries. We prove
that the method itself naturally predicts the correct assingment of
the electric charge spectrum along with the relation between the gauge
couplings of the groups involved therein for each particular model
in this class. 


\end{abstract}

\section{Introduction}

The general method of exactly solving models with high symmetries
- based on the gauge group \textbf{$SU(3)_{c}\otimes SU(n)_{L}\otimes U(1)_{Y}$}
that undergoes a spontaneous symmetry breakdown (SSB) in its electro-weak
sector - was proposed several years ago by Cot\u{a}escu \cite{key-1}
and applied by the author \cite{key-2} - \cite{key-7} to the so
called 3-3-1 models. This exact algebraical approach is employed here
to prove that, in the case of the \textbf{$SU(3)_{c}\otimes SU(4)_{L}\otimes U(1)_{Y}$}
(3-4-1) models \cite{key-8} - \cite{key-18}, the correct electric
charge assignment for all the fermion representations (and thus for
all the bosons) can be predicted by just setting the versors $\nu_{i}$
in the general Weinberg transformation (gWt). The gWt is designed
to bring the massive vector fields from the gauge basis to the their
physical basis through a $SO(n-1)$ rotation in the parameter space.
At the same time, the procedure by itself naturally ensures at this
stage the correct gauge coupling matching without resorting to any
supplemental hypothesis. 

The paper is organized as follows: Sec. 2 reviews the main aspects
of the general method, Sec. 3 presents its predictions regarding the
lectric charge spectrum and gauge couplings matching in each particular
case of versor setting for the \textbf{$SU(4)_{L}\otimes U(1)_{Y}$}
model, while Sec.4 sketches our conclusions.

\section{The general method - a brief review}

All the details of constructing a consistent algebraical approach
designed to exactly solve gauge models with high symmetries can be
found in Ref. \cite{key-1}. We restrict ourselves here to outline
only the main results concerning the electric charge operator and
close related topics. The electric charge assignment must ensure such
fermion representations that all the anomalies cancel by an interplay
between generations.

\subsection{Irreducible representations of $SU(n)_{L}\otimes U(1)_{Y}$ }

We focus on the fermion representations of the $SU(n)_{L}\otimes U(1)_{Y}$
electro-weak model. The main piece is the group $SU(n)$ and its two
fundamental irreducible unitary representations (irreps) $\mathbf{n}$
and $\mathbf{n^{*}}$ which give different classes of tensors of ranks
$(r,s)$ as direct products like $(\otimes{\textbf{n}})^{r}\otimes(\otimes{\textbf{n}}^{*})^{s}$.
These tensors have $r$ lower and $s$ upper indices for which we
reserve the notation, $i,j,k,\cdots=1,\cdots,n$. As usually, we denote
the irrep $\rho$ of $SU(n)$ by indicating its dimension, ${\mathbf{n}}_{\rho}$.
The $su(n)$ algebra can be parameterized in different ways, but here
it is convenient to use the hybrid basis of Ref. \cite{key-1} consisting
of $n-1$ diagonal generators of the Cartan subalgebra, $D_{\hat{i}}$,
labeled by indices $\hat{i},\hat{j},...$ ranging from $1$ to $n-1$,
and the generators $E_{j}^{i}=H_{j}^{i}/\sqrt{2}$, $i\not=j$, related
to the off-diagonal real generators $H_{j}^{i}$ \cite{key-19,key-20}.
This way the elements $\xi=D_{\hat{i}}\xi^{\hat{i}}+E_{j}^{i}\xi_{i}^{j}\in su(n)$
are now parameterized by $n-1$ real parameters, $\xi^{\hat{i}}$,
and by $n(n-1)/2$ $c$-number ones, $\xi_{j}^{i}=(\xi_{i}^{j})^{*}$,
for $i\not=j$. The advantage of this choice is that the parameters
$\xi_{j}^{i}$ can be directly associated to the $c$-number gauge
fields due to the factor $1/\sqrt{2}$ which gives their correct normalization.
In addition, this basis exhibit good trace orthogonality properties,
\begin{equation}
Tr(D_{\hat{i}}D_{\hat{j}})=\frac{1}{2}\delta_{\hat{i}\hat{j}},\quad Tr(D_{\hat{i}}E_{j}^{i})=0\,,\quad Tr(E_{j}^{i}E_{l}^{k})=\frac{1}{2}\delta_{l}^{i}\delta_{j}^{k}\,.\label{Eq.1}\end{equation}
 When we consider different irreps, $\rho$ of the $su(n)$ algebra
we denote $\xi^{\rho}=\rho(\xi)$ for each $\xi\in su(n)$ such that
the corresponding basis-generators of the irrep $\rho$ are $D_{\hat{i}}^{\rho}=\rho(D_{\hat{i}})$
and $E_{j}^{\rho\, i}=\rho(E_{j}^{i})$.

\subsection{Fermion sector}

The $U(1)_{Y}$ transformations are nothing else but phase factor
multiplications. Therefore - since the coupling constants $g$ for
$SU(n)_{L}$ and $g^{\prime}$ for the $U(1)_{Y}$ are assinged -
the transformation of the fermion tensor $L^{\rho}$ with respect
to the gauge group of the theory reads \begin{equation}
L^{\rho}\rightarrow U(\xi^{0},\xi)L^{\rho}=e^{-i(g\xi^{\rho}+g^{\prime}y_{ch}\xi^{0})}L^{\rho}\label{Eq.2}\end{equation}
 where $\xi=\in su(n)$ and $y_{ch}$ is the chiral hypercharge defining
the irrep of the $U(1)_{Y}$ group parametrized by $\xi^{0}$. For
simplicity, the general method deals with the character $y=y_{ch}g^{\prime}/g$
instead of the chiral hypercharge $y_{ch}$, but this mathematical
artifice does not affect in any way the results. Therefore, the irreps
of the whole gauge group $SU(n)_{L}\otimes U(1)_{Y}$ are uniquely
detemined by indicating the dimension of the $SU(n)$ tensor and its
character $y$ as $\rho=({\textbf{n}}_{\rho},y_{\rho})$.

\subsection{Gauge Fields}

As in all kinds of gauge theories, the interactions are mediated by
gauge fields that are vector fields (massive or massless) that couple
different fermion fields in a particular manner, by introducing the
so called covariant derivatives. The gauge fields are in our notation
$A_{\mu}^{0}=(A_{\mu}^{0})^{*}$ and $A_{\mu}=A_{\mu}^{+}\in su(n)$
respectively, while the needed covariant derivatives are defined as
$D_{\mu}L^{\rho}=\partial_{\mu}L^{\rho}-ig(A_{\mu}^{a}T_{a}^{\rho}+y_{\rho}A_{\mu}^{0}I^{\rho})L^{\rho}$
where $T_{a}^{\rho}$ are generators (let them be diagonal or off-diagonal)
of the $su(n)$ algebra.

\subsection{Minimal Higgs Mechanism}

The scalar sector, organized as the so called minimal Higgs mechanism
(mHm), is flexible enough to produce the SSB in one step up to the
$U(1)_{em}$ symmetry and, consequently, generate masses for the plethora
of particles and bosons in the model. The scalar sector consists of
$n$ Higgs multiplets $\phi^{(1)}$, $\phi^{(2)}$, ... $\phi^{(n)}$
satisfying the orthogonality condition $\phi^{(i)+}\phi^{(j)}=\phi^{2}\delta_{ij}$
in order to eliminate the unwanted Goldstone bosons that could survive
the SSB. $\phi$ is a gauge-invariant real scalar field while the
Higgs multiplets $\phi^{(i)}$ transform according to the irreps $({\textbf{n}},y^{(i)})$
whose characters $y^{(i)}$ are arbitrary numbers that can be organized
into the diagonal matrix $Y={\textrm{diag}}\left(y^{(1)},y^{(2)},\cdots,y^{(n)}\right)$.
In addition, the Higgs sector needs, in our approach, a parameter
matrix $\eta={\textrm{diag}}\left(\eta{}^{(1)},\eta{}^{(2)},...,\eta{}^{(n)}\right)$with
the property ${\textrm{Tr}}(\eta^{2})=1-\eta_{0}^{2}$ in order to
supply a non-degenerate boson mass spectrum after SSB took place.
The scalar potential is assumed to have an absolute minimum for $\phi=\langle\phi\rangle\not=0$
that is, $\phi=\langle\phi\rangle+\sigma$ where $\sigma$ is the
unique surviving physical Higgs field. Therefore, one can always define
the unitary gauge where the Higgs multiplets, $\hat{\phi}^{(i)}$
have the components $\hat{\phi}_{k}^{(i)}=\delta_{ik}\phi=\delta_{ik}(\langle\phi\rangle+\sigma)$

\subsection{Electric and neutral charges}

The charge spectrum of the model is close related to the problem of
finding the basis of the physical neutral bosons. First of all, the
method ensures th separation of the electromagnetic potential $A_{\mu}^{em}$
corresponding to the surviving $U(1)_{em}$ symmetry. The one-dimensional
subspace of the parameters $\xi^{em}$ associated to this symmetry
assumes a particular direction in the parameter space $\lbrace\xi^{0},\xi^{\hat{i}}\rbrace$
of the whole Cartan subalgebra. This is uniquely determined by the
$n-1$ - dimensional unit vector $\nu$ and the angle $\theta$ giving
the subspace equations $\xi^{0}=\xi^{em}\cos\theta$ and $\xi^{\hat{i}}=\nu_{\hat{i}}\xi^{em}\sin\theta$.
On the other hand, since the Higgs multiplets in unitary gauge are
invariant under $U(1)_{em}$ transformations, one remains with the
condition $D_{\hat{i}}\xi^{\hat{i}}+Y\xi^{0}=0$ which yields $Y=-D_{\hat{i}}\nu^{\hat{i}}\tan\theta\equiv-(D\cdot\nu)\tan\theta$.
In other words, the new parameters $(\nu,\theta)$ determine all the
characters $y^{(i)}$ of the irreps of the Higgs multiplets and hence
these will be considered the principal parameters of the model. Therefore
we one deal with $\theta$ and $\nu$ (which has $n-2$ independent
components) instead of $n-1$ parameters $y^{(i)}$. Under these circumstances,
one can easily compute the mass term of the gauge bosons depending
on the parameters $\theta$ and $\nu_{\hat{i}}$. Evidently, $A_{\mu}^{em}$
does not appear in the mass term, so it remains massless. The other
neutral gauge fields ${A'}_{\mu}^{\hat{i}}$ have the non-diagonal
mass matrix (Eq,(53) in Ref. \cite{key-1}). This can be brought in
diagonal form with the help of a $SO(n-1)$ transformation, $A_{\mu}^{'\hat{i}}=\omega_{\cdot\;\hat{j}}^{\hat{i}\;\cdot}Z_{\mu}^{\hat{j}}$
, which leads to the physical neutral bosons $Z_{\mu}^{\hat{i}}$
with well-defined masses. Performing this $SO(n-1)$ transformation
the physical neutral bosons are completely determined. The transformation
\begin{eqnarray}
A_{\mu}^{0} & = & A_{\mu}^{em}\cos\theta-\nu_{\hat{i}}\omega_{\cdot\;\hat{j}}^{\hat{i}\;\cdot}Z_{\mu}^{\hat{j}}\sin\theta,\nonumber \\
A_{\mu}^{\hat{k}} & = & \nu^{\hat{k}}A_{\mu}^{em}\sin\theta+\left(\delta_{\hat{i}}^{\hat{k}}-\nu^{\hat{k}}\nu_{\hat{i}}(1-\cos\theta)\right)\omega_{\cdot\;\hat{j}}^{\hat{i}\;\cdot}Z_{\mu}^{\hat{j}}.\label{Eq.3}\end{eqnarray}
 which switches from the original diagonal gauge fields, $(A_{\mu}^{0},A_{\mu}^{\hat{i}})$
to the physical ones, $(A_{\mu}^{em},Z_{\mu}^{\hat{i}})$. This is
called the generalized Weinberg transformation (gWt).

Now one can identify the charges of the particles with the coupling
coefficients of the currents with respect to the above determined
physical bosons. Thus, we find that the spinor multiplet $L^{\rho}$
(of the irrep $\rho$) has the following electric charge matrix \begin{equation}
Q^{\rho}=g\left[(D^{\rho}\cdot\nu)\sin\theta+y_{\rho}\cos\theta\right],\label{Eq.4}\end{equation}
 and the $n-1$ neutral charge matrices \begin{equation}
Q^{\rho}(Z^{\hat{i}})=g\left[D_{\hat{k}}^{\rho}-\nu_{\hat{k}}(D^{\rho}\cdot\nu)(1-\cos\theta)-y_{\rho}\nu_{\hat{k}}\sin\theta\right]\omega_{\cdot\;\hat{i}}^{\hat{k}\;\cdot}\label{Eq.5}\end{equation}
 corresponding to the $n-1$ neutral physical fields, $Z_{\mu}^{\hat{i}}$.
All the other gauge fields, namely the charged bosons $A_{j\mu}^{i}$,
have the same coupling, $g/\sqrt{2}$, to the fermion multiplets.

\section{$SU(4)_{L}\otimes U(1)_{Y}$ models}

The general method must be based on the following assumptions in order
to give viable results when it is applied to concrete models:

({\small I}) the spinor sector must be put (at least partially) in
pure left form using the charge conjugation (see for details Appendix
B in Ref. \cite{key-1})

({\small II}) the minimal Higgs mechanism must be employed with its
arbitrary parameters $(\eta_{0},\eta)$ satisfying the condition ${\textrm{Tr}}(\eta^{2})=1-\eta_{0}^{2}$
and giving rise to traditional Yukawa couplings in unitary gauge

({\small III}) the coupling constant, $g$, is the same with the first
one of the SM

({\small IV}) at least one $Z$-like boson should satisfy the mass
condition $m_{Z}=m_{W}/\cos\theta_{W}$ established in the SM and
experimentally confirmed.

Conditions (II) and (IV) lead to a realistic non-degenerate mass spectrum
for particular classes of the 3-4-1 model that will be presentetd
elsewhere \cite{key-21}. For our purpose here condition (III) plays
a crucial role.

In the following, we will use the standard generators $T_{a}$ of
the $su(4)$ algebra. Therefore, as the Hermitian diagonal generators
of the Cartan subalgebra one deals, in order, with $D_{1}=T_{3}=\frac{1}{2}{\textrm{diag}}(1,-1,0,0)$,
$D_{2}=T_{8}=\frac{1}{2\sqrt{3}}\,{\textrm{diag}}(1,1,-2,0)$, and
$D_{3}=T_{15}=\frac{1}{2\sqrt{6}}\,{\textrm{diag}}(1,1,1,-3)$ respectively.
At the same time, we denote the irreps of the electroweak model under
consideration here by $\rho=({\textbf{n}}_{\rho},y_{ch}^{\rho})$
indicating the genuine chiral hypercharge $y_{ch}$ instead of $y$.
Therefore, the multiplets - subject to anomaly cancellation - of the
3-4-1 model of interest here will be denoted by $({\textbf{n}}_{color},{\textbf{n}}_{\rho},y_{ch}^{\rho})$. 

There are three distinct cases leading to a discrimination among models
of the 3-4-1 class, according to their electric charge assignment.
They are: (i) versors $\nu_{1}=1$, $\nu_{2}=0$, $\nu_{3}=0$, (ii)
versors $\nu_{1}=0$, $\nu_{2}=1$, $\nu_{3}=0$, and (iii) versors
$\nu_{1}=0$, $\nu_{2}=0$, $\nu_{3}=-1$, respectively. At the same
time, one assumes the condition $e=g\sin\theta_{W}$ established in
the SM.

\subsection{Case 1 (versors $\nu_{1}=1$, $\nu_{2}=0$, $\nu_{3}=0$)}

In this case, the lepton 4-plet obeys the fundamental irrep of the
gauge group $\rho=(\mathbf{4},0)$ . Eq. (\ref{Eq.4}) yields:

\begin{equation}
Q^{(4,0)}=eT_{3}^{(4)}\frac{\sin\theta}{\sin\theta_{W}},\label{Eq.6}\end{equation}
which leads to the lepton 4-plet $\left(\begin{array}{cccc}
e_{\alpha}^{c}, & e_{\alpha}, & \nu_{\alpha}, & N_{\alpha}\end{array}\right)_{L}^{T}\sim(\mathbf{4},0)$ if and only if $\sin\theta=2\sin\theta_{W}$ holds. 

For the two families ($i=1,2$) of quarks transforming in the same
way under the gauge group $\left(\begin{array}{cccc}
J_{i}, & u_{i}, & d_{i}, & D_{i}\end{array}\right)_{L}^{T}\sim(\mathbf{4^{*}},-1/3)$ and for the third one that transforms as $\left(\begin{array}{cccc}
J_{3}, & d_{3}, & u_{3}, & U_{3}\end{array}\right)_{L}^{T}\sim(\mathbf{4},+2/3)$, the electric charge operator will take, respectively, the forms 

\begin{equation}
Q^{(4^{*},-\frac{1}{3})}=e\left[T_{3}^{(4^{*})}\frac{\sin\theta}{\sin\theta_{W}}-\frac{1}{3}\left(\frac{g^{\prime}}{g}\right)\frac{\cos\theta}{\sin\theta_{W}}\right],\label{Eq.7}\end{equation}

\begin{equation}
Q^{(4,+\frac{2}{3})}=e\left[T_{3}^{(4)}\frac{\sin\theta}{\sin\theta_{W}}+\frac{2}{3}\left(\frac{g^{\prime}}{g}\right)\frac{\cos\theta}{\sin\theta_{W}}\right],\label{Eq.8}\end{equation}
compatible with the known quark charges if and only if

\begin{equation}
\frac{g^{\prime}}{g}=\frac{\sin\theta_{W}}{\sqrt{1-4\sin^{2}\theta_{W}}}.\label{Eq.9}\end{equation}

For the sake of completness we show the fermion representations of
this class of models.

\textbf{Lepton families}\begin{equation}
\begin{array}{ccccc}
f_{\alpha L}=\left(\begin{array}{c}
e_{\alpha}^{c}\\
e_{\alpha}\\
\nu_{\alpha}\\
N_{\alpha}\end{array}\right)_{L}\sim(\mathbf{1,4},0)\end{array}\label{Eq.10}\end{equation}

\textbf{Quark families}\begin{equation}
\begin{array}{ccc}
Q_{iL}=\left(\begin{array}{c}
J_{i}\\
u_{i}\\
d_{i}\\
D_{i}\end{array}\right)_{L}\sim(\mathbf{3,4^{*}},-1/3) &  & Q_{3L}=\left(\begin{array}{c}
J_{3}\\
-d_{3}\\
u_{3}\\
U_{3}\end{array}\right)_{L}\sim(\mathbf{3},\mathbf{4},+2/3)\end{array}\label{Eq.11}\end{equation}
\begin{equation}
\begin{array}{c}
(d_{3L})^{c},(d_{iL})^{c},(D_{iL})^{c},\sim(\mathbf{3},\mathbf{1},+1/3)\end{array}\label{Eq.12}\end{equation}

\begin{equation}
(u_{3L})^{c},(u_{iL})^{c},(U_{3L})^{c}\sim(\mathbf{3},\mathbf{1},-2/3)\label{Eq.13}\end{equation}

\begin{equation}
\begin{array}{ccccc}
(J_{3L})^{c}\sim(\mathbf{3,1},-5/3) &  &  &  & (J_{iL})^{c}\sim(\mathbf{3,1},+4/3)\end{array}\label{Eq.14}\end{equation}
with $\alpha=1,2,3$ and $i=1,2$. 

In the representations presented above one can assume, like in majority
of the papers in the literature, that the third generation of quarks
transforms differently from the other two ones. This could explain
the unusual heavy masses of the third generation of quarks, and especially
the uncommon properties of the top quark. The capital letters $J$
denote the exotic quarks included in each family. They exhibit electric
charges $\pm4/3$ and $\pm5/3$. 

This possible choice of the versors $\nu_{1}=1$, $\nu_{2}=0$, $\nu_{3}=0$
has led us to the very class of 3-4-1 models with exotic electric
charges \cite{key-8} - \cite{key-13} whose phenomenology predicted
by our method will be \emph{in extenso} analysed in Ref. \cite{key-21}.

\subsection{Case 2 (versors $\nu_{1}=0$, $\nu_{2}=1$, $\nu_{3}=0$)}

Due to $T_{8}=\frac{1}{2\sqrt{3}}\,{\textrm{diag}}(1,1,-2,0)$ there
is no room for a plausible electric charge operator assigning only
$0$ and $\pm e$ in the lepton 4-plets. Therefore, this case is ruled
out as long as one does not allow for exotic electric charges like,
for instance $\pm2e$, in the lepton sector.

\subsection{Case 3 (versors $\nu_{1}=0$, $\nu_{2}=0$, $\nu_{3}=-1$) }

In this case, no 4-plet obeys the fundamental irrep of the gauge group
$\rho=(\mathbf{4},0)$. Notwithstanding, since for the lepton 4-plet
one can assign two different chiral hypercharges $-\frac{1}{4}$ and
$-\frac{3}{4}$ respectively, we get two sub-cases leading to two
different versions of the class of 3-4-1 models without exotic electric
charges. The coupling matching, as we will see in the following, assumes
the same relation in both sub-cases.

From Eq. (\ref{Eq.4}), it is straightforward that the lepton family
exhibits the electric charge operator

\begin{equation}
Q^{(4^{*},-\frac{1}{4})}=e\left[-T_{15}^{(4^{*})}\frac{\sin\theta}{\sin\theta_{W}}-\frac{1}{4}\left(\frac{g^{\prime}}{g}\right)\frac{\cos\theta}{\sin\theta_{W}}\right],\label{Eq.15}\end{equation}
for the first sub-case. This leads to the lepton representation $\left(\begin{array}{cccc}
e_{\alpha}, & \nu_{\alpha}, & N_{\alpha}, & N_{\alpha}^{\prime}\end{array}\right)_{L}^{T}\sim(\mathbf{4^{*}},-\frac{1}{4})$ including two new kinds of neutral leptons ($N_{\alpha}$, $N_{\alpha}^{\prime}$). 

In the second subcase, the electric charge operator will be represented
as 

\begin{equation}
Q^{(4,-\frac{1}{4})}=e\left[-T_{15}^{(4)}\frac{\sin\theta}{\sin\theta_{W}}-\frac{3}{4}\left(\frac{g^{\prime}}{g}\right)\frac{\cos\theta}{\sin\theta_{W}}\right],\label{Eq.16}\end{equation}
 leading to the lepton families $\left(\begin{array}{cccc}
\nu_{\alpha}, & e_{\alpha}^{-}, & E_{\alpha}^{-}, & E_{\alpha}^{\prime-}\end{array}\right)_{L}^{T}\sim(\mathbf{4},-\frac{3}{4})$ that allow for new charged leptons ($E_{\alpha}^{-}$, $E_{\alpha}^{\prime-}$).

After a little algebra, both Eqs (\ref{Eq.15}) and (\ref{Eq.16})
require - via the compulsory condition $\sin\theta=\sqrt{\frac{3}{2}}\sin\theta_{W}$,
since the unique allowed electric charges in the lepton sector are
$0$ and $\pm e$ - the coupling matching:

\begin{equation}
\frac{g^{\prime}}{g}=\frac{\sin\theta_{W}}{\sqrt{1-\frac{3}{2}\sin^{2}\theta_{W}}}.\label{Eq.17}\end{equation}

Once these assingments are assumed, the quarks will aquire their electric
charges from the following operators

\begin{equation}
Q^{(4^{*},\frac{5}{12})}=e\left[-T_{15}^{(4^{*})}\frac{\sin\theta}{\sin\theta_{W}}+\frac{5}{12}\left(\frac{g^{\prime}}{g}\right)\frac{\cos\theta}{\sin\theta_{W}}\right]\label{Eq.18}\end{equation}

\begin{equation}
Q^{(4,-\frac{1}{12})}=e\left[-T_{15}^{(4)}\frac{\sin\theta}{\sin\theta_{W}}-\frac{1}{12}\left(\frac{g^{\prime}}{g}\right)\frac{\cos\theta}{\sin\theta_{W}}\right]\label{Eq.19}\end{equation}

\subsubsection{Case 3a }

With the first of the above mentioned assumptions, the fermion representations
are:

\textbf{Lepton families}\begin{equation}
\begin{array}{ccccc}
f_{\alpha L}=\left(\begin{array}{c}
e_{\alpha}\\
\nu_{\alpha}\\
N_{\alpha}\\
N_{\alpha}^{\prime}\end{array}\right)_{L}\sim(\mathbf{1,4^{*}},-\frac{1}{4}) &  &  &  & \left(e_{\alpha L}\right)^{c}\sim(\mathbf{1,1},1)\end{array}\label{Eq.20}\end{equation}

\textbf{Quark families}\begin{equation}
\begin{array}{ccc}
Q_{iL}=\left(\begin{array}{c}
u_{i}\\
d_{i}\\
D_{i}\\
D_{i}^{\prime}\end{array}\right)_{L}\sim(\mathbf{3,4},-1/12) &  & Q_{3L}=\left(\begin{array}{c}
-d_{3}\\
u_{3}\\
U\\
U^{\prime}\end{array}\right)_{L}\sim(\mathbf{3},\mathbf{4^{*}},5/12)\end{array}\label{Eq.21}\end{equation}
\begin{equation}
\begin{array}{c}
(d_{3L})^{c},(d_{iL})^{c},(D_{iL})^{c},(D_{iL}^{\prime})^{c}\sim(\mathbf{3},\mathbf{1},+1/3)\end{array}\label{Eq.22}\end{equation}

\begin{equation}
(u_{3L})^{c},(u_{iL})^{c},(U_{L})^{c},(U_{L}^{\prime})^{c}\sim(\mathbf{3},\mathbf{1},-2/3)\label{Eq.23}\end{equation}

with $\alpha=1,2,3$ and $i=1,2$. We recovered the same fermion content
as the one of the model presented in Refs. \cite{key-14,key-18}.

\subsubsection{Case 3b }

With the second of the above mentioned assumptions, the fermion representations
are:

\textbf{Lepton families}\begin{equation}
\begin{array}{ccccc}
f_{\alpha L}=\left(\begin{array}{c}
\nu_{\alpha}\\
e_{\alpha}^{-}\\
E_{\alpha}^{-}\\
E_{\alpha}^{\prime-}\end{array}\right)_{L}\sim(\mathbf{1,4},-3/4) &  &  &  & (e_{\alpha L})^{c},(E_{\alpha L})^{c},(E_{\alpha L}^{\prime})^{c}\sim(\mathbf{1,1},1)\end{array}\label{Eq.24}\end{equation}

\textbf{Quark families}\begin{equation}
\begin{array}{ccc}
Q_{iL}=\left(\begin{array}{c}
d_{i}\\
-u_{i}\\
U_{i}\\
U_{i}^{\prime}\end{array}\right)_{L}\sim(\mathbf{3,4^{*}},5/12) &  & Q_{3L}=\left(\begin{array}{c}
u_{3}\\
d_{3}\\
D\\
D^{\prime}\end{array}\right)_{L}\sim(\mathbf{3},\mathbf{4},-1/12)\end{array}\label{Eq.25}\end{equation}
\begin{equation}
\begin{array}{c}
(d_{3L})^{c},(d_{iL})^{c},(D_{L})^{c},(D_{L}^{\prime})^{c}\sim(\mathbf{3},\mathbf{1},+1/3)\end{array}\label{Eq.26}\end{equation}

\begin{equation}
(u_{3L})^{c},(u_{iL})^{c},(U_{iL})^{c},(U_{iL}^{\prime})^{c}\sim(\mathbf{3},\mathbf{1},-2/3)\label{Eq.27}\end{equation}
with $\alpha=1,2,3$ and $i=1,2$. We recovered the same fermion content
as the one of the model presented in Refs. \cite{key-17,key-18}.

With this assignment the fermion families (in each of the above displayed
cases) cancel the axial anomalies by just an interplay between them,
although each family remains anomalous by itself. Thus, the renormalization
criteria are fulfilled and the method is validated once more from
this point of view. Note that one can add at any time sterile neutrinos
- \emph{i.e.} right-handed neutrinos $\nu_{\alpha R}\sim(\mathbf{1,1},0)$
- that could pair in the neutrino sector of the Ld with left-handed
ones in order to eventually generate tiny Dirac or Majorana masses
by means of an adequate see-saw mechanism. These sterile neutrinos
do not affect anyhow the anomaly cancelation, since all their charges
are zero. Moreover, their number is not restricted by the number of
flavors in the model

\section{Concluding remarks}

In this brief report we have obtained the correct electric charge
assignment and matching the gauge couplings for some particular anomaly-free
models of the 3-4-1 class, by just using the prescriptions of the
general method of exactly solving gauge models with high symmetries.
All the results are simply consequences of a proper versor choice
in the general Weinberg transformation. This approach represents a
complementary way to discriminate among different particular 3-4-1
models, in addition to the well-known classification \cite{key-16}
based on the parameters $b$ and $c$. However our approach is a little
bit more restrictive, since it leaves out the model investigated in
Ref. \cite{key-15} which can be reproduced by none versor setting
in our method.

The complex phenomenology of the above obtained 3-4-1 models - such
as boson mass spectrum, neutrino masses, extra-neutral bosons and
neutral currents, bileptons etc. - will be investigated in a future
work.

\end{document}